\input epsf
\documentstyle[preprint,aps]{revtex}
\input epsf

\begin{document}
\draft
%\begin{flushright}
%UTAS-PHYS-94-27
%\end{flushright}
\title{Estimates for Forward-backward Asymmetry in
	 $B\rightarrow K^*(892)\; l^+ l^-$}
\author{Dongsheng Liu}
\address{Department of Physics, University of Tasmania\\
Hobart, AUSTRALIA 7001}
\date{Sept., 1994}
\maketitle

\begin{abstract}

Forward-backward charge asymmetry in rare dilepton
$B$-decays is formulated with the
assumption of an on-shell b-quark. We find the asymmetry is
expressed in terms of two spatially transverse helicity amplitudes,
which are determined by combining the data of
$D\rightarrow K^*(892)\; l^+ \nu$ with heavy flavour symmetry.
We estimate the charge asymmetry of $B\rightarrow K^*(892)\; l^+ l^-$
for large energy leptons against various top masses in the standard model.
\end{abstract}


%\date{#1} for dating receipt at Editorial Offices of APS
\section{Introduction}
Rare $B$-decays have been the focus of many experimental
and theoretical considerations\cite{xsphoton,rare}.
This is due to the amount of information concerning the standard model (SM)
that can be extracted from these processes.
The rare decays proceed through flavour
changing neutral current (FCNC) vertices which are absent at the tree level
and thus provide a good probe of the SM at the quantum (loop) level.  On
the other hand, rare $B$-decays are sensitive to quark mixing angles
$V_{td}, V_{ts}$ and $ V_{tb}$,
hence their determination will yield valuable information
on CKM matrix elements and consequently shed some light on CP violation
in the SM.  These processes are also dependent on the top quark
mass $m_t$ through the dominant internal top quark line so that
a comparison between theoretical estimates as a
function of $m_t$ and experiment may lead to constraints on
the top quark mass.
In this letter we focus on the rare decays $B \rightarrow K^* l^+l^-
(l = e, \mu)$.

The motivation for this work comes from the observation
that for rare decays $b \rightarrow s\; l^+l^-$
(which proceed through $\gamma$, Z and $W^\pm$ exchange diagrams)
forward-backward charge asymmetry of dilepton production
has the potential to be fairly large in the SM \cite{AMM}.
For $m_t/M_W \geq 2$,
as suggested by the CDF value of $m_t=174\pm 17$~GeV,
the contribution of the $Z$-exchange
diagrams becomes important, leading to a substantial asymmetry.
Such an asymmetry in inclusive $B \rightarrow X_s\; l^+l^-$ processes
has been recently studied in the SM and some non-standard models \cite{AGM}.
Consequently, we feel that similar investigations should be carried out
on {\em exclusive} rare $B$-decays. In particular, we hope to examine the
sensitivity of forward-backward asymmetry to the top mass and extensions
of the SM.
A main source of uncertainty in all such studies has been
the evaluation of hadronic matrix elements (HME) for specific exclusive
decays.  Such evaluations involve long range nonperturbative QCD
effects which render the calculations model dependent.
During the past few years there has been considerable progress in formulating
HME for cases in which the
mesons contain a heavy quark\cite{hqet,BSmb,DL}.  For
example, if the c-quark is treated heavy along with the b-quark, all
the form factors parameterizing
the HME for heavy to light $0^- \rightarrow 0^-$ and $0^- \rightarrow 1^-$
processes can be written in terms of a set of six universal form factors
which represent the underlying QCD dynamics \cite{IW42,DD}.
These form factors carry the heavy flavour symmetry and are
independent of quark current operators.
Thus they permit the model independent study of {\em exclusive}
$B$ decays into light flavour mesons in the small recoil region.
In the rare dilepton $B$-decays, it is necessary to restrict oneself
to a dilepton invariant mass well below the $J/\psi$-mass
or above the $\psi'$-mass to avoid the resonance background.
As far as the $B\rightarrow K^* l^+ l^-$ channel is concerned,
the forward-backward asymmetry expressed in terms of
spatially transverse form factors is suppressed as one tends toward
smaller dilepton masses (see Eq.\ (5)).
In this letter, we shall concentrate on
regions close to the maximum value of the dilepton mass .
We shall firstly formulate forward-backward
charge asymmetry in rare dilepton $B$-decays.
Secondly, using the
assumption of an on-shell b-quark, we demonstrate that the asymmetry is
determined by two helicity amplitudes $h_\pm$. Further,
combining the data of $D\rightarrow K^*(892)\; l^+ \nu$ for individual
form factors with
heavy flavour symmetry, we estimate the asymmetry of
$B\rightarrow K^*(892)\; l^+ l^-$
for several top masses. Finally, we present discussions
concerning our numerical results and comment roughly
on the potential signatures of new physics in rare dilepton
$B$-decays.

\section{Formulae of forward-backward asymmetry}
Let us begin with an effective Hamiltonian relevant to flavour-changing
one-loop processes
$b\rightarrow s l^+ l^-$  \cite{dilep}
\begin{equation}
H_{eff}=\frac{G_F}{\sqrt{2}}\left( \frac \alpha {4\pi s_W^2}\right) [\bar s%
\Gamma _\mu ^Ab\;\bar l\gamma ^\mu (1-\gamma _5)l+\bar s\Gamma _\mu ^Bb\;%
\bar l\gamma ^\mu (1+\gamma _5)l],
\end{equation}
with effective vertices
$$
\Gamma _\mu ^{A(B)}=A(B)\gamma _\mu (1-\gamma _5)-im_bs_W^2F_2\sigma _{\mu
\nu }q^\nu (1+\gamma _5)/q^2.
$$
In this Hamiltonian, heavy particles, $W^{\pm }$-bosons and the top quark are
integrated out and their masses together with QCD corrections are absorbed
into coefficient functions,
\begin{equation}
A(B,F_2)=\displaystyle\sum_{q=u,c,t}V_{qs}^{*}V_{qb}A_q(B_q,F_2^q),
\end{equation}
which are dominated in the SM by the top quark contributions,
except for the long
distance effect which proceeds mainly through CKM favored $c\bar c$
intermediate vector meson states. Here $V_{qs}$ and $V_{qb}$ are elements of
the CKM matrix and $A_q(B_q,F_2^q)$ are given in ref.~\cite{dilep,QCD,long}.
The effective quark current
$\bar s\Gamma _\mu ^{A(B)}b$ in question has two different structures;
the parametrization for the matrx element of $V-A$ currents in terms of
invariant form factors is
\begin{equation}\begin{array}{rl}
\langle p, \phi|\bar s\gamma ^\mu (1-\gamma _5)b |P\rangle = &
[a_{+}(q^2)(P+p)_\mu +a_{-}(q^2)(P-p)_\mu ]P^\nu \phi_\nu ^{*} \\
&+f(q^2)\phi_\mu ^{*}
 +ig(q^2)\epsilon _{\mu\nu\alpha\beta}\phi^{*\nu }P^\alpha p^\beta.
\end{array}\end{equation}
In analogy to this we have for the magnetic-moment operator
\begin{equation}\begin{array}{rl}
-\displaystyle\frac{i}{q^2}
\langle p, \phi|\bar s\sigma_{\mu \nu }q^\nu (1+\gamma _5)b |P\rangle = &
[\tilde a_{+}(q^2)(P+p)_\mu +\tilde a_{-}(q^2)(P-p)_\mu ]P^\nu \phi_\nu ^{*} \\
& +\tilde f(q^2)\phi_\mu ^{*}
  +i\tilde g(q^2)\epsilon _{\mu\nu\alpha\beta}\phi^{*\nu }P^\alpha p^\beta,
\end{array}\end{equation}
along with a condition of current conservation,
$$(M^2-m^2)\tilde a_+ + q^2 \tilde a_- + \tilde f =0.$$
Here $M\;(P_\mu)$ and $m\;(p_\mu)$ are masses (momenta)
of parent and daughter mesons, respectively,
$\phi_\mu$ the polarization vector of daughter mesons (satisfying
$\phi_\mu p^\mu =0$) and $q=P-p$ the momentum transfer into the dilepton.

The differential forward-backward asymmetry  of the $l^+$
that we are to study is defined by
$$d\Gamma_{FB}(q^2)=
  \int^1_0\; d\Gamma(\cos\theta_l)-\int^0_{-1}\; d\Gamma(\cos\theta_l),
$$
in which $\pi-\theta_l$ is the polar angle of the $l^+$
with respect to the direction of motion of the decaying meson in
the $l^+l^-$-frame. It can be derived from the decay
distribution in ref. \cite{DD}
\begin{equation}\displaystyle
\frac{d\Gamma_{FB}}{dq^2}
= \displaystyle\frac{1}{64 \pi^3}
  \left ( \frac{G_F}{\sqrt{2}} \right )^2 \mid V\mid^2 \frac{\lambda\;q^2}{M^2}
[ (\mid H_+^L\mid^2 -\mid H_-^L\mid^2 )
- (\mid H_+^R\mid^2 -\mid H_-^R\mid^2)],
\end{equation}
where $\lambda=\sqrt{(v\cdot p)^2-m^2}$ is a measure of the recoil
and turns out to be the magnitude of the momentum of the daughter meson
in the parent rest frame with
$v\cdot p=\displaystyle\frac{1}{2M}(M^2+m^2-q^2)$.
As long as the u-quark is ignored, one has
$\mid V\mid=\displaystyle\frac{\alpha}{4\pi s^2_W}\mid V^*_{cs}V_{cb}\mid$
from unitarity of the CKM matrix.
The helicity amplitudes appearing here are defined as
\begin{equation}
H_\pm^{L(R)}=\displaystyle\sum_{q=c,t}
		   A_q(B_q) h_\pm + m_bs_W^2F_2^q \tilde h_\pm,
\end{equation}
where $h_\pm\equiv f\pm\lambda M g$
and $\tilde h_\pm\equiv\tilde f\pm\lambda M \tilde g$.
In Eq.\ (5) we set the limit of $m_l=0$ and therefore
there is no mixture between left- and right-handed leptons.
The small lepton mass only manifests itself at the lower boundary
$q^2=4m_l^2$ logarithmically, but since we
are only interested in the region near the upper boundary,
we may safely use such a limit.
Assuming the $B$-meson contains an on-shell b-quark, one finds
\begin{equation}
\tilde h_\pm=\frac{1}{q^2}(M-v\cdot p\mp \lambda)\; h_\pm,
\end{equation}
(see the Appendix for detailed dicussion).
This leads us into helicity amplitudes
\begin{equation}
H_\pm^L=(A+m_bs_W^2F_2\frac{M-v\cdot p\mp \lambda}{q^2})\; h_\pm,
\end{equation}
\begin{equation}
H_\pm^R=(B+m_bs_W^2F_2\frac{M-v\cdot p\mp \lambda}{q^2})\; h_\pm.
\end{equation}
For comparision, we also present the transverse distribution
of the decay rate
\begin{equation}\displaystyle
\frac{d\Gamma_T}{dq^2}
= \displaystyle\frac{1}{48 \pi^3}
  \left ( \frac{G_F}{\sqrt{2}} \right )^2 |V|^2 \frac{\lambda\;q^2}{M^2}
[ (\mid H_+^L\mid^2 +\mid H_-^L\mid^2 )
+ (\mid H_+^R\mid^2 +\mid H_-^R\mid^2 )].
\end{equation}
We emphasize that all forms presented here also hold
for $B$-decays into higher $K$-resonances \cite{DD}.

\section{Helicity form factors and numerical results}
Isgur and Wise \cite{IW42} have suggested using heavy flavour symmetry to
relate the form factors for $B\rightarrow K^* l^+l^-$-decays with
those for semileptonic $D$-decays having the same meson, i.e. $K^*(892)$
in the final state. However, in the $e$ and $\mu$-channels of $D$-decays
it is difficult to measure the $V$-$A$ form factor $a_-$ as
it is suppressed by $m_l^2/q^2$.
Consequently, it is hard for one to obtain the form factor
$\tilde a_+$ (or $\tilde a_-$) in the rare $B$-decays using this method.
But we are indeed allowed to determine those form factors occuring in
the asymmetry when we concentrate on the small recoil region.
Burdman and Donoghue have instead used $SU(3)$
light flavour symmetry to relate rare $B$-decays with semileptonic
$B$-decays\cite{BD}. In this letter we will use the approach of Isgur
and Wise and the form factors $A_1$ and $V$
(proportional to $f$ and $g$, respectively) of
$D\rightarrow K^*(892)\; l^+ \nu$
processes measured by E691, E687 and CLEO groups \cite{DK1}.
All of these groups assume the nearest pole
dominance for the $q^2$-dependence of form factors.
Since the range of $q^2$ in this decay is only about $1$~GeV$^2$
(small compared to heavy pole masses),
the resulting form factors are not sensitive to
the parameterization of the $q^2$-dependence.

In the leading order of heavy quark effctive theory, form factors scale as
\begin{equation}\displaystyle
 h_\pm
=\left [\frac{\alpha_s(m_b)}{\alpha_s(m_c)}\right ]^{-6/25}
  \sqrt{\frac{M}{M_D}}\;h^D_\pm.
\end{equation}
In this equation helicity amplitudes are evaluated at the same
$v\cdot p$, namely that $h_\pm$ at $q^2=M^2+m^2-2M v\cdot p$
are related to $h^D_\pm$ at
$q_D^2=M_D^2+m^2-2M_D v\cdot p$.
The region in which heavy quark symmetries can be applied
to weak decays may be
determined by the average momentum transfer $Q_l$ of the light degrees
of freedom. It is estimated heuristically for $D\rightarrow K^*$ decay
by \cite{Isgur} that
$$Q_l^2 < \frac{m_q}{m} \frac{m_q}{M_D}(M_D-m)^2\simeq (0.13 M_D)^2,$$
with light quark mass $m_q=330$~MeV.
Thus we expect the heavy flavour symmetry to be a reliable approximation
in the whole
physical range of $q_D^2=[0,\; (M_D-m)^2]$, corresponding to
$q^2=[q_0^2,\; (M-m)^2]$ in $B\rightarrow K^*$ modes.
Here the lower limit  $q_0^2=(4.07$~GeV$)^2$ serves as
an experimental cut-off for the avaliable data and restricts us to
a range that is away from the peaks of $J/\psi$ and $\psi'$ and
above the $c\bar c$ threshold. Thus decay distributions
are dominated by the top quark plus the continuum  $c\bar c$
involved via four-qaurk operators.
With the coefficients given in ref.~\cite{QCD} and
the parameters listed below, we integrate forward-backward asymmetry,
along with the transverse decay rate, in the momentum interval
of $[q_0^2,\; (M-m)^2]$. The numerical results for representative top masses
are shown in Table 1. Like in ref.\ \cite{ALiu},
both the asymmetry and decay rate are normalized to
$\Gamma_0=\Gamma(B\rightarrow X_s\;J/\psi)Br(J/\psi\rightarrow l^+\; l^-)$,
using the experimental input
$Br(B\rightarrow X_s\;J/\psi)=(1.09\pm0.04)\%$\cite{cleoI}
and $Br(J/\psi\rightarrow l^+\; l^-)=(5.98\pm 0.25)\%$\cite{pdg}.

Although we believe that reliable estimates for the asymmetry
are provided in the stated kinematic region,
we need an alternative prescription
to determine helicity amplitudes. This will enable us to probe
decay distributions below $q^2_0$ for the sake of increasing
the statistics. In this situation Eq.\ (11)
is only used to gain $h_\pm$ at a $q^2_{m}$ which is approaching to,
but not identical with, the maximum value of $q^2=(M-m)^2$
where $h_+=h_-$.
It seems appropriate in the small recoil region of
$B\rightarrow K^*$ decays to approximate
the $q^2$-dependence of form factors by the nearest pole dominance
\begin{equation}
 \left ( 1-\displaystyle\frac{q^2}{M'^2}\right )\; h_\pm (q^2)
=\left ( 1-\displaystyle\frac{q^2_{m}}{M'^2}\right )\; h_\pm (q^2_{m}),
\end{equation}
as the end-point is close to the pole. Here $M'$ is instead a pole mass
of the current involving b- and s-quarks and
we will make no distinction between the vector and axial masses.
As a check, we first evaluate the integrated asymmetry with
helicity amplitudes in Eq.\ (13) over the interval of $[q_0^2,\; (M-m)^2]$.
The numerical results give a difference within $5\%$ ($7\%$ for
transverse decay rate), from that with Eq.\ (11). This provides
us confidence to
extrapolate the data of $D$-decays to the region below $q_0^2$
but above the $D\bar D$ continuum threshold
$q_1^2=(3.74$~GeV$)^2$. Table 2 is devoted
to numerical integrals over $[q_1^2,\; (M-m)^2]$.
The contributions to decay distributions of individual
helicity amplitudes are illustrated in Fig. 1 for the left-handed lepton
and in Fig. 2 for the right-handed lepton .

The parameters that appear in our calculations are taken as
(in GeV for masses)
$$ M=5.28, \quad M_D=1.87, \quad m=0.892,\quad M(D_s^*)=2.11,
\quad M(D_{s1})=2.54,
$$
$$s_W^2=0.233, \quad
\mid V_{ts}^{*}V_{tb}\mid =\mid V_{cs}^{*}V_{cb}\mid=0.044,\quad
 m_b=4.9,\quad m_c=1.5,\quad M'=5.42.
$$

\section{Discussions}

We shall now interprete the numerical results qualitatively.
Firstly we recall that it is well confirmed in semileptonic
$B\rightarrow D^*$ and $D\rightarrow K^*$ that
$\mid\!\! h_-\!\!\mid > \mid\!\! h_+\!\!\mid$.
An analogous relationship that the (transverse) negative helicity
of the $K^*$ dominates over the positive one, i.e.
$\mid\!\! H^{L(R)}_-\!\!\mid > \mid\!\! H^{L(R)}_+\!\!\mid$
is found in the regions we have considered for dilepton rare decays.
This is because the s-quark produced by the effective hamiltonian
in Eq.\ (1) is of
predominantly negative (or left-handed) helicity. Upon hadronization,
the s-quark picks up a spectator quark having both helicities
with equal probability
and thus forms a $K^*$ in favour of the negative
one. Actually, the contribution of $H^{L(R)}_-$
differs dramatically from that of $H^{L(R)}_+$
due to the large mass of the $B$-meson.
Secondly with no four-quark operator mixing in,
we have the SM values of the coefficient functions in Eq.\ (2)
\begin{equation}
A_t=1.94, \quad B_t=-0.136, \quad s^2_W F_2^t=-0.141,
\end{equation}
for the effective vertices at $m_t=174$~GeV. The QCD corrections
are  partly responsible
for the small coupling of the right-handed lepton. Obviously,
left-handed leptons dominate the final state. This feature is
common to all top mass cases listed in Table 1.
As the four-quark operator
$
\bar s\gamma_\mu (1-\gamma _5)b\;\bar c\gamma ^\mu (1-\gamma _5)c
$
enters through a vector current and shifts both $A$ and $B$
by about 0.1, there is very little change in the chirality
pattern. We should remind ourselves that the mixing of other
four-quark operators is even smaller\cite{QCD,misiak},
so the chirality of leptons represented by Eq.\ (13) strongly
influences the resulting decay distribution. As a final word, we realize
that $H_+^R$ suffers from both helicity and QCD suppression
and is thus negligble, $H_+^L$ is largely suppressed by helicity
and $H_-^R$ is only reduced by the QCD correction.
Hence, the chirality of both quarks and leptons determined by the SM
remains clearly manifested in exclusive processes; the forward-backward
charge asymmetry is negative, sensitive to the magnitude of $H_-^L$.
With a relative small effect for right-handed leptons,
the asymmetry does not differ from the transverse decay rate too much.
Given a longitudinal helicity amplitude which is not particularly
large, the decay distribution in the regime near
the zero recoil comes mainly from transverse helicity amplitudes
due to a $q^2$ enhancement.
In other words, the magnitude of the asymmetry is expected to be
comparable with the decay rate, to which we refer as a large
aymmetry of lepton production predicted by the SM in
$B\rightarrow K^*\; l^+l^-$ processes.

To obtain more precise results, we must account for
the deviation from the heavy flavour symmetry due to finite masses
of b- and c-quarks.
The leading corrections to the ratio of form factors in
$B\rightarrow \rho\;l\bar\nu$ to that in $D\rightarrow \rho\;l\bar\nu$
are evaluated at the end-point in the framework of a constituent
quark model \cite{Dib}.
It is found that the deviation from the limit of $m_Q\rightarrow\infty$
is encouragingly small, being of order $15\%$. With this indication
we assume modification of a similar size for the
$B\rightarrow K^*\; l^+l^-$ decay. Such an analysis will
be reported elsewhere.

We conclude with a number of comments on potential signals of
new physics beyond the SM in rare dilepton $B$-decays. Recent measurments
of $B\rightarrow X_s\gamma$ by the CLEO group \cite{xsphoton} result in
a constraint on the magitude of $F_2$\cite{AGM}
$$0.44 \geq |F_2| \leq 0.60,$$
which is close to the value of the SM. The sign of $F_2$ is irrelevant to
radiative processes, but is important in dilepton decays because of
interference. In the SM, $F_2$ is negative relative to $A$
and contrastly new physics
may manifest itself if a positive one of similar size was allowed \cite{susy}.
Such an $F_2$ would produce
an enhancement of about factor 1.5 in the asymmetry and the decay rate.
Meanwhile, any different input of both the magnitude and the sign
for $A$ and $B$ from that of the SM will lead to visible changes to
the forward-backward asymmetry as well as the decay rate
in rare dilepton decays.
It is, however, interesting to describe a special possibility in which
the importance of left-handed leptons is replaced by right-handed
leptons. While it is possible for the decay rate not to be sensitive to
the change, the large negative asymmetry will be turned into a positive one.
In such a case, it is the measurment of the asymmetry that tells us about
new physics.
Alternatively, if the asymmetry were small, our estimates should fail
to be reliable, as corrections from the quark mass would show up when
different contributions in Eq.\ (5) would tend to cancel each other.
Fortunately, this is not the case in the SM. Nevertheless, a small
asymmetry itself, if measurements find it to be so, can be regarded
as a signature of new physics.
We are optimistic that current and future $B$-physics facilities
will provide data on rare dilepton decays so that we may
test the SM
and probe new physics in this sensitive territory.

\acknowledgements

The author thanks R. Delbourgo for many discussions and encouragement
during the work along with
P. Jarvis and N. Jones for their help in preparation of this letter.
He is also grateful to the Australian Research Council for their
financial support, under grant number A69231484.

\appendix{}

\section*{Relations of Form Factors}
When a heavy meson contains an on-shell heavy quark
the matrix element for the heavy to light
transition takes the form \cite{BSmb,DL,DD}
\begin{equation}
\displaystyle
<p, \phi|\bar q\Gamma h_v|P\!=\!Mv>=
\sqrt{M}{\rm Tr}\left[
[(G_1 + G_2\not\! p) v\cdot\phi + (G_3+G_4\not\! p )\not\!\phi ]
\Gamma \frac{1+\not v}2\gamma_5\right],
\end{equation}
with $\Gamma$ an arbitrary matrix in Dirac space. The invariant overlap
integrals $G_i$ are functions of $v\cdot p$ and bear the heavy
quark symmetry.
Comparing Eq.\ (3) and Eq.\ (4) in the main text to traces with
$\Gamma=\gamma _\mu (1-\gamma _5)$ and
$\sigma _{\mu\nu }q^\nu (1+\gamma _5)$
gives
\begin{equation}
\left ( \begin{array}{c}
	  f \\
	  g \end{array}
\right )=-2\sqrt{M}
\left ( \begin{array}{cc}
	  1 & \quad -v\cdot p \\
	  0         & \quad \displaystyle\frac{1}{M}
	  \end{array}
\right )
\left ( \begin{array}{c}
	  G_3 \\
	  G_4 \end{array}
\right ),
\end{equation}
and
\begin{equation}
\left ( \begin{array}{c}
	 \tilde f \\
	 \tilde g \end{array}
\right )=\frac{2\sqrt{M}}{q^2}
\left ( \begin{array}{cc}
	  -(M-v\cdot p)              & \quad M(v\cdot p)-m^2\\
	   \displaystyle\frac{1}{M}  & \quad -1
	  \end{array}
\right )
\left ( \begin{array}{c}
	  G_3 \\
	  G_4 \end{array}
\right ),
\end{equation}
respectively. Then removing $G_3$ and $G_4$ we find
\begin{equation}
\left ( \begin{array}{c}
	  \tilde f \\
	  \lambda M\tilde g \end{array}
\right )=\frac{1}{q^2}
\left ( \begin{array}{cc}
	  M-v\cdot p &    \quad -\lambda  \\
	   -\lambda  & \quad M-v\cdot p
	  \end{array}
\right )
\left ( \begin{array}{c}
	  f \\
	  \lambda M g \end{array}
\right ),
\end{equation}
agreeing with that of ref. \cite{IW42}.
Futhermore, transforming into transverse helicity amplitudes,
$h_\pm=f\pm\lambda M g$ \cite{KS} and
$\tilde h_\pm=\tilde f\pm\lambda M \tilde g$,
diagonalizes
the relation with eigenvalues $M-v\cdot p\mp\lambda$. This leaves
\begin{equation}
\tilde h_\pm=\frac{1}{q^2}(M-v\cdot p\mp \lambda)\; h_\pm.
\end{equation}

\newpage
{\large Table captions}:

{\bf Table 1} The integrated asymmetry and transverse decay rate
		    (in $10^{-4}$)
		    over the region of $q^2=[q_0^2,\; (M-m)^2]$
		    with helicity amplitudes
		    determined by the data of $D\rightarrow K^*(892)\; l^+ \nu$
		    (see Eq.\ (11) of the main text).
		    All entries are normalized to
		    $\Gamma_0=\Gamma(B\rightarrow X_s\;J/\psi)
		    Br(J/\psi\rightarrow l^+\; l^-)$.

{\bf Table 2} The asymmetry and transverse decay rate (in $10^{-4}$)
		   integrated over the region of $q^2=[q_1^2,\; (M-m)^2]$
		   with helicity amplitudes
		   determined by combining the data of $D\rightarrow
		   K^*(892)\; l^+ \nu$ with nearest pole domination
		   (see Eq.\ (13) of the main text). The normalization
		   is the same as Table 1.
\vskip 2cm
{\large Figure captions}:

{\bf Figure 1} Decay distributions
$\displaystyle\frac{1}{\Gamma_0}\frac{d\Gamma_{\pm}}{ds}\times 10^4$,
			for the left-handed lepton in terms of
			the negative (dot-dashed line) and positive (solid line)
			helicity amplitudes, for which Eq.\ (13) is used.
			We define that $s=q^2/M^2$ and vertical dashed lines
			at $s=0.488,\; 0.501, \; 0.594$ indicate
			$\psi'$-peak, $D\bar D$-threshold and $q_D^2=0$,
			respectively. The top mass is taken as $m_t=174$~GeV.

{\bf Figure 2} The same as Fig. 1 for the right-handed lepton.
\newpage
\vskip 3cm
\begin{table}
\begin{tabular}{c|c|c|c||c|c|c}
    \multicolumn{1}{|c|}{Top mass (GeV)}
 &  \multicolumn{1}{c|}{$\quad\Gamma_{FB}^{(L)}\quad$}
 &  \multicolumn{1}{c|}{$\quad\Gamma_{FB}^{(R)}\quad$}
 &  \multicolumn{1}{c||}{$\quad\Gamma_{FB}\quad$}
 &  \multicolumn{1}{c|}{$\quad\Gamma_{T}^{(L)}\quad$}
 &  \multicolumn{1}{c|}{$\quad\Gamma_{T}^{(R)}\quad$}
 &  \multicolumn{1}{c|}{$\quad\Gamma_{T}\quad$}   \cr
\hline
    \multicolumn{1}{|c|}{131}
&   \multicolumn{1}{c|}{-0.213}
&   \multicolumn{1}{c|}{-0.002}
&   \multicolumn{1}{c||}{-0.211}
&   \multicolumn{1}{c|}{0.374}
&   \multicolumn{1}{c|}{0.003}
&   \multicolumn{1}{c|}{0.377} \cr
\hline
    \multicolumn{1}{|c|}{150}
&   \multicolumn{1}{c|}{-0.288}
&   \multicolumn{1}{c|}{-0.015}
&   \multicolumn{1}{c||}{-0.273}
&   \multicolumn{1}{c|}{0.506}
&   \multicolumn{1}{c|}{0.025}
&   \multicolumn{1}{c|}{0.531} \cr
\hline
    \multicolumn{1}{|c|}{174}
&   \multicolumn{1}{c|}{-0.378}
&   \multicolumn{1}{c|}{-0.008}
&   \multicolumn{1}{c||}{-0.359}
&   \multicolumn{1}{c|}{0.644}
&   \multicolumn{1}{c|}{0.013}
&   \multicolumn{1}{c|}{0.658} \cr
\hline
    \multicolumn{1}{|c|}{200}
&   \multicolumn{1}{c|}{-0.501}
&   \multicolumn{1}{c|}{-0.026}
&   \multicolumn{1}{c||}{-0.475}
&   \multicolumn{1}{c|}{0.877}
&   \multicolumn{1}{c|}{0.043}
&   \multicolumn{1}{c|}{0.920} \cr
\end{tabular}
\end{table}
\begin{center}
Table 1.
\end{center}

\vskip 3cm
\begin{table}
\begin{tabular}{c|c|c|c||c|c|c}
    \multicolumn{1}{|c|}{Top mass (GeV)}
 &  \multicolumn{1}{c|}{$\quad\Gamma_{FB}^{(L)}\quad$}
 &  \multicolumn{1}{c|}{$\quad\Gamma_{FB}^{(R)}\quad$}
 &  \multicolumn{1}{c||}{$\quad\Gamma_{FB}\quad$}
 &  \multicolumn{1}{c|}{$\quad\Gamma_{T}^{(L)}\quad$}
 &  \multicolumn{1}{c|}{$\quad\Gamma_{T}^{(R)}\quad$}
 &  \multicolumn{1}{c|}{$\quad\Gamma_{T}\quad$}   \cr
\hline
    \multicolumn{1}{|c|}{131}
&   \multicolumn{1}{c|}{-0.578}
&   \multicolumn{1}{c|}{-0.004}
&   \multicolumn{1}{c||}{-0.573}
&   \multicolumn{1}{c|}{0.857}
&   \multicolumn{1}{c|}{0.007}
&   \multicolumn{1}{c|}{0.863} \cr
\hline
    \multicolumn{1}{|c|}{150}
&   \multicolumn{1}{c|}{-0.748}
&   \multicolumn{1}{c|}{-0.006}
&   \multicolumn{1}{c||}{-0.742}
&   \multicolumn{1}{c|}{1.11}
&   \multicolumn{1}{c|}{0.01}
&   \multicolumn{1}{c|}{1.12} \cr
\hline
    \multicolumn{1}{|c|}{174}
&   \multicolumn{1}{c|}{-0.999}
&   \multicolumn{1}{c|}{-0.025}
&   \multicolumn{1}{c||}{-0.975}
&   \multicolumn{1}{c|}{1.48}
&   \multicolumn{1}{c|}{0.04}
&   \multicolumn{1}{c|}{1.52} \cr
\hline
    \multicolumn{1}{|c|}{200}
&   \multicolumn{1}{c|}{-1.36}
&   \multicolumn{1}{c|}{-0.08}
&   \multicolumn{1}{c||}{-1.29}
&   \multicolumn{1}{c|}{2.02}
&   \multicolumn{1}{c|}{0.11}
&   \multicolumn{1}{c|}{2.12} \cr
\end{tabular}
\end{table}
\begin{center}
Table 2.
\end{center}
%\newpage
%\epsfxsize = 16cm \epsfbox{fig1.ps}
%\newpage
%\epsfxsize = 16cm \epsfbox{fig2.ps}

\end{document}